# Demystifying Graviton Detection

Igor Pikovski[1,2]

[1]*Department of Physics, Stevens Institute of Technology, Hoboken, New Jersey 07030, USA*
[2]*Department of Physics, Stockholm University, Stockholm, Sweden*
*pikovski@stevens.edu*

Detecting single gravitons is possible through the extraction of discrete energy quanta from passing gravitational waves. For decades, this was considered hopeless, with estimates suggesting a detector the size of Jupiter. Here, I explain how these earlier estimates relied on the highly inefficient electron ionization process, while the absorption through the full mass quadrupole of a macroscopic system can bring graviton detection within experimental reach. Operating in the LIGO frequency band solves the challenges of graviton sources and noise suppression, allowing gravitons to be extracted from known passing gravitational waves and detection events to be heralded by classical gravitational-wave observations. Combined with the ability to resolve single energy quanta in macroscopic systems, such as in quantum acoustics, this opens a realistic path toward graviton detection. I also discuss the general limitations of particle detection in establishing an underlying quantum theory, illustrated by an energy-nonconserving semiclassical alternative. Nevertheless, the semiclassical alternative is not standard classical wave physics, and extracting $E = \hbar\omega$ from a passing wave constitutes bona fide particle detection. This opens the way not only to the first detection of individual gravitons, but also to experimentally probing linearized quantum gravity and the graviton counting statistics of gravitational waves.

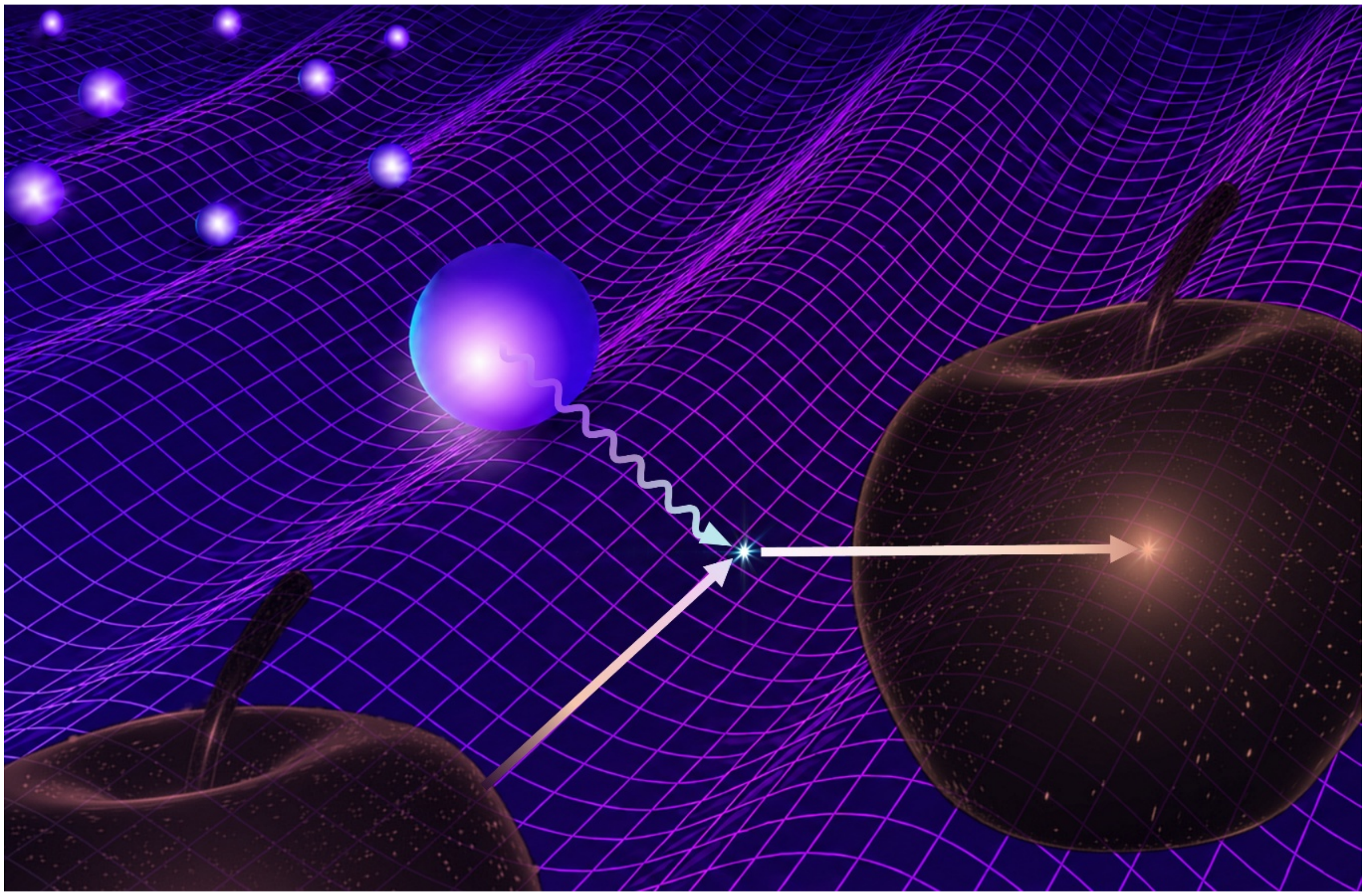

## Introduction

The graviton has long occupied a strange place in physics: central to a quantum description of gravity, yet seemingly forever beyond experiment. If gravity obeys quantum mechanics, then the gravitational interaction should be mediated by a quantum particle, the graviton, analogous to the particles that mediate the other fundamental forces. Mathematically, they are described by linearized quantum gravity (Box 1). But the weak coupling of gravity seemed to prevent experimental access to gravitons. Steven Weinberg [1], for example, calculated the spontaneous emission of a graviton from an atom and found that the process would take far longer than the age of the universe. Freeman Dyson later examined several possible detection methods [2], showing in one example that a LIGO-like detector would have to be so massive it would collapse into a black hole.

Another alternative, more closely related to how other particles are routinely detected, was considered by Dyson [2], as well as by Rothman and Boughn [3]: a gravitational analogue of the photoelectric effect, where a graviton is absorbed from a gravitational-wave background through electronic transitions in an atom. But this method also appeared hopeless, requiring a detector the size of Jupiter in close proximity to a neutron star. Even before confronting the enormous challenge of background noise, these results (and common sense) suggested that graviton detection lay hopelessly beyond practical reach.

Yet a new approach has shown that graviton detection can be achieved [4]. It is based on the same premise of energy absorption as in the photoelectric analogy above, but focuses on a far more efficient interaction process. Inspired by quantum optics, the solution combines a first-principles analysis of graviton-matter interactions with the ability to detect individual quanta of energy in macroscopic systems. With a clearer perspective and with experimental capabilities that were out of reach even a few years ago, detecting a single graviton no longer seems fundamentally impossible. Instead, single-graviton detection now appears to be a realistic experimental goal.

This result shows how single graviton detection can be achieved – a still difficult but now feasible task. At the same time, successful graviton detection must be understood in the right context. It would not provide final proof of a quantum theory of gravity. That is because particle detection can also be interpreted through a semi-classical model, if one is willing to abandon energy conservation. But this is no longer standard classical wave physics. The ability to extract $E = \hbar\omega$ from a passing wave is bona fide particle detection and would provide direct access to the single-particle regime of the interaction between gravity and matter, opening the door to broader experimental exploration of the predictions of quantum gravity and its possible alternatives.

**Box 1: Linearized Quantum Gravity**

Gravitons are a near-universal prediction of quantum gravity, because they already arise from quantizing the linearized limit of general relativity. Rather than working with the full metric space-time $g_{\mu\nu}(x)$, whose curvature describes gravity in Einstein's theory, one assumes that gravity is sufficiently weak so that $g_{\mu\nu}(x) \approx \eta_{\mu\nu} + h_{\mu\nu}(x)$, where $\eta_{\mu\nu}$ is the flat Minkowski metric of special relativity, and $h_{\mu\nu}(x)$ is the weak gravitational field. Keeping only terms linear in $h_{\mu\nu}$, general relativity looks like an ordinary field theory on flat spacetime: $h_{\mu\nu}(x)$ propagates much like the fields of the Standard Model, and its equations closely resemble Maxwell's equations.

Despite its simplicity, this approximation already reproduces much of the observed phenomenology of general relativity [1], including gravitational waves. Most importantly, it can be quantized similarly to electromagnetism, as Matvei Bronstein [5] first showed in his 1935 doctoral thesis "Quantization of Gravitational Waves" (tragically, Bronstein fell victim to Stalin's purges soon after). This introduces gravitational particles now called "gravitons," in close analogy with photons. The theory reproduces the quantum version of Einstein's quadrupole formula for gravitational radiation (see Box 2), as well as the Newtonian interaction between masses through the exchange of gravitons. These results capture the two complementary roles of the graviton: as a virtual particle mediating gravitational forces, and as a real "on-shell" particle forming the indivisible energy quanta of gravitational waves. Each graviton carries energy

$$E = \hbar\omega,$$

where $\hbar$ is Planck's reduced constant and $\omega$ the angular frequency of the wave, directly analogous to a photon. Its interaction with matter is captured by the Hamiltonian density $H = -\frac{1}{2}T^{\mu\nu}h_{\mu\nu}$, where $T^{\mu\nu}$ is the stress-energy tensor. This mirrors electromagnetism where the charge current $j^{\mu}$ couples to the field $A_{\mu}$ through $H_{\mathrm{EM}} = j^{\mu}A_{\mu}$, except that gravity couples through a two-index spin-2 field and to a "charge" that is mass and energy, which are always positive.

Linearized quantum gravity is a mathematically consistent framework. It breaks down only once gravity becomes sufficiently strong that the gravitational field interacts nonlinearly with itself. It is at this point that the deeper problems of quantum gravity arise. Nevertheless, any successful theory of quantum gravity is expected to reduce to linearized quantum gravity in the weak-field limit, and thus to contain gravitons.

## Catching gravitons

The key to single-graviton detection turns out to be the resonant exchange of energy between a gravitational wave and matter. It can be described by the cross-section

$$\sigma \sim \frac{absorbed\ power}{incident\ flux}.$$

Dyson [2], Rothman and Boughn [3] found it to be vanishingly small for atomic transitions:

$$\sigma \sim L_p^2 \approx 10^{-70} m^2,$$

roughly the Planck area. Thus an atom, whose physical area is roughly the Bohr radius squared $a_0^2 \approx 10^{-20} m^2$, is essentially transparent to gravitons. This Planck-scale cross-section, a direct reflection of the weakness of gravity, is often quoted as the fundamental obstacle to detecting gravitons.

Yet this conclusion can be circumvented. The Planck-scale result applies to a very specific process: gravito-ionization, in which an electron absorbs a single graviton from a passing gravitational wave and changes its orbital (see Figure 1a). The crucial insight is that this is not the only possible process to consider, nor is it the most suitable one. Instead of an atom, what if the detector were a much larger mass, while still undergoing measurable quantum jumps in energy? In fact, massive systems deep in the quantum regime have recently been proposed for conventional gravitational wave detection, such as superfluid helium resonators [6] as modern variants of the resonant bars pioneered by Joe Weber [7]. The core new idea was that in such detectors, individual quantum levels could in principle be measured, much like transitions in an atom, but involving a vastly larger mass. This possibility arises from recent developments in quantum physics: the ability to bring increasingly macroscopic systems into the quantum regime [8]. And that's precisely where the long-standing intuition about graviton detection is overturned.

Building on the above reasoning, my group and I worked out the idea quantitatively. One can compute the exact absorption and emission probabilities using the Hamiltonian of linearized quantum gravity [4]. But the essential scaling can already be understood from a simple argument, summarized in Box 2. It shows that the Planck-area cross-section is not fundamental. Rather, the more general result is

$$\sigma \sim \frac{G\omega}{c^3} Q,$$

where $Q$ is the detector mass-quadrupole which scales with mass and size of the system.

For microscopic systems this cross-section is tiny. An electron in an atom has $Q \sim m_e a_0^2$, which reproduces $\sigma \sim L_p^2$ (the Bohr-Sommerfeld quantization results in $\omega Q = \hbar$). But this describes only how an *electron* absorbs energy from the gravitational wave. Physically, that is like trying to generate electricity from gravitational waves. Instead of electrons, one can consider other transitions in systems with vastly larger mass quadrupoles. A resonant bar of size $L$ and mass $M$, for example, yields $\sigma \sim G\omega M L^2 / c^3$. In this case, the energy is not absorbed by an electron, but by a *phonon*. The cross-section for this process, graviton-to-phonon conversion, can therefore be larger by even forty orders of magnitude. This changes the picture completely, bringing a possible graviton detector down from astronomical dimensions to human scales (Figure 1b).

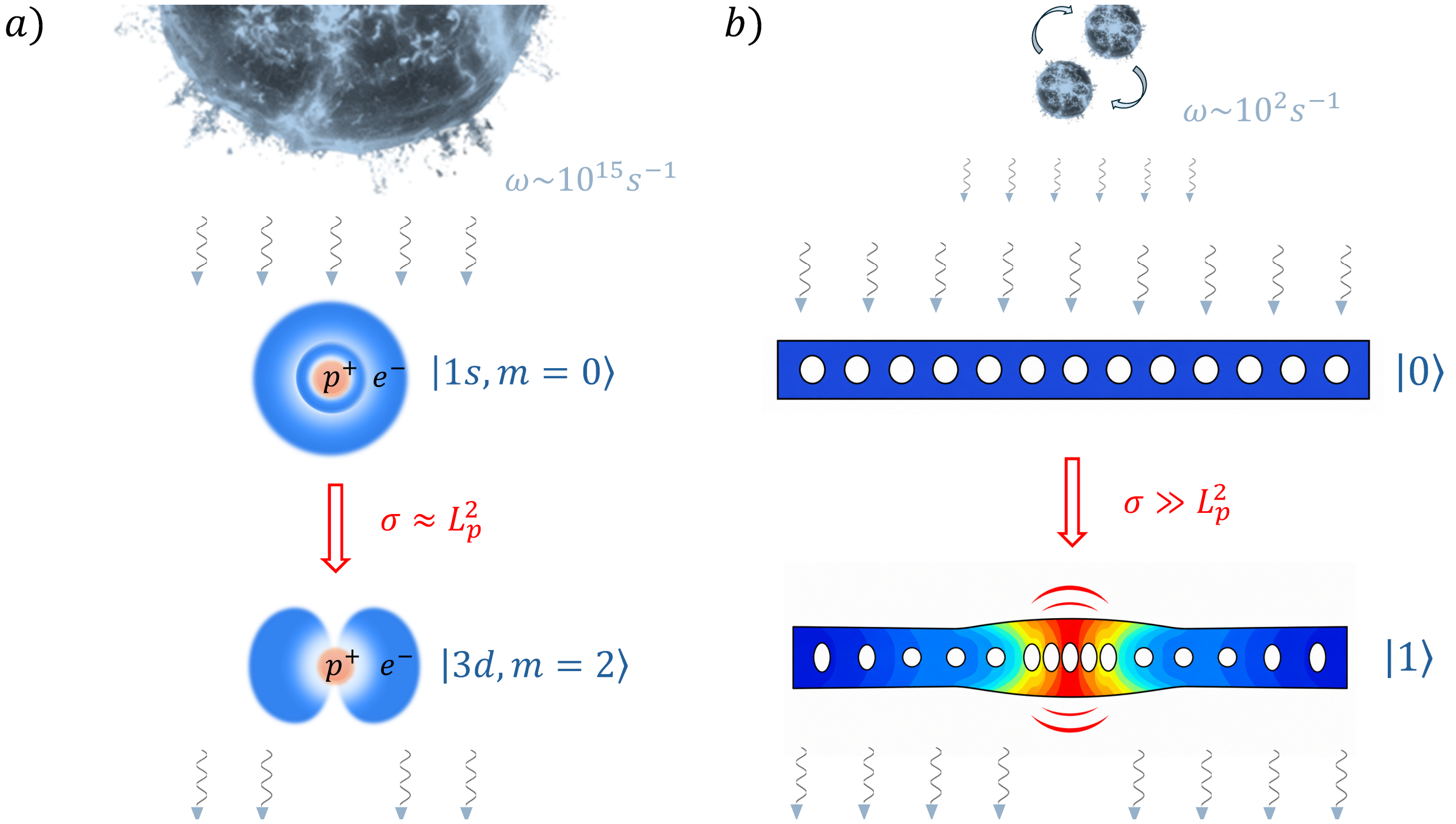

*Figure 1: From atomic absorption to macroscopic graviton detection. Graviton detection was thought to be impossible, but access to quantized macroscopic systems now makes it an experimental target. a) Atomic absorption illustrates the traditional difficulty of graviton detection [1-3]: a graviton can be absorbed by an electronic quadrupole transition from 1s to 3d orbitals, but the excitation probability is vanishingly small as the graviton cross-section is on the order of the Planck area. It also requires graviton sources that are resonant with the high-frequency transition, for example from thermal emission from a star with the exponentially suppressed high-frequency Wien-tail of the spectrum. These requirements make detection appear hopeless. b) Absorption in a ground-state-cooled macroscopic acoustic resonator changes the picture [4]: graviton-to-phonon conversion has a much larger cross-section because of the vastly larger mass quadrupole compared with electronic transitions. The graviton flux from passing gravitational waves can also be enormous in the relevant frequency range, for example from compact binary mergers detected by LIGO. A single graviton is detected when the resonator registers a change of exactly one quantum of energy,* $E = \hbar\omega$, *analogous to a quantum jump in an atom. Background noise is mitigated by cross-correlating the event with the conventional detection of the passing gravitational wave. Extracting a single graviton from a passing wave thus becomes possible with realistic technology.*

**Box 2: The graviton cross-section**

Einstein's quadrupole formula gives the power emitted in gravitational waves as

$$P \sim G\dddot{Q}^2/c^5,$$

where $G$ is Newton's constant, $c$ the speed of light and $\dddot{Q}$ the third time-derivative of the mass quadrupole of the emitter. A changing quadrupole thus emits gravitational waves. But the same logic can be turned around to predict how a detector would absorb energy instead, an insight that led Joe Weber in 1960 to propose the first gravitational wave detector [7]. Importantly, the mass quadrupole scales with the mass and size of the system as $Q{\sim}mx^2$, and it changes through the strains caused by the gravitational wave. For single graviton detection, only a single quantum of energy is absorbed, captured by a transition matrix element $\langle f|Q|i\rangle$ in which the strain in the detector changes on the order of $\sqrt{\hbar/m\omega}$. For a monochromatic wave resonant with the detector, the power associated with a single graviton transition becomes $P{\sim}G\omega^6|\langle f|Q|i\rangle|^2/c^5 \sim G\hbar\omega^5 Q/c^5$. The flux associated with a single graviton is $F \sim \hbar\omega \cdot \omega^3/c^2$, corresponding to one quantum of energy spread over roughly one wavelength during one oscillation period. Together, this yields the absorption cross-section:

$$\sigma = P/F \sim G\omega Q/c^3.$$

This simple but general expression captures the key to graviton detection.

**Quantum sensing in macro-systems**

The ability for macroscopic matter to absorb gravitational wave energy was of course well known for classical detection [1,7]. The novelty is to consider and measure energy absorption in discrete steps, the same way one observes a transition in a quantized atom. Single graviton absorption or emission is then no longer confined to the microscopic realm, and benefits from the much larger gravitational "charge", the mass quadrupole. But detecting a single graviton requires more than measuring the (continuous) displacement resulting from a gravitational wave, as in conventional gravitational-wave detection. It requires measuring energy itself, and resolving a change by a single quantum, $E = \hbar\omega$.

In recent years, exactly this ability has begun to emerge through increasing quantum control over macroscopic systems. One example is the so-called field of "quantum acoustics". The phonon concept – the quanta of sound – has been essential to solid state physics for over a century. But the ability to create, manipulate and detect states of individual phonons is very recent, building on advancements in the fields of opto- and electro-mechanics [9]. Quantum acoustic systems come in many forms: piezoelectric resonators, phononic crystals, surface-acoustic waves, bulk-acoustic wave resonators, Brillouin optomechanics, or sound in superfluid Helium, to name a few. The phonons are controlled and measured either by swapping them onto another quantum system, such as a superconducting qubit or an optical photon, or through direct energy-coupling that enables so-called quantum non-demolition measurements (Figure 2). And the ability to count individual phonons has recently been demonstrated in a range of mesoscopic systems.

Much of the progress in this field has been motivated by the broader effort to develop quantum technologies, such as for frequency transduction, quantum sensing or quantum information processing. But the quantum control at macroscopic scales is also profoundly affecting fundamental physics. Measuring individual quanta in macroscopic systems, such as in quantum acoustics, turns graviton detection from an apparent impossibility into a task for quantum engineering. Still, this task remains difficult. Existing systems are still too small to be relevant for graviton absorption and would need to be scaled up in mass, roughly from the microgram range to gram-scale systems. The energy resolution has also to be improved to detect quanta at around kilohertz frequencies. But the tools are developing rapidly and these goals may well be achievable in the near future with acoustic or other systems.

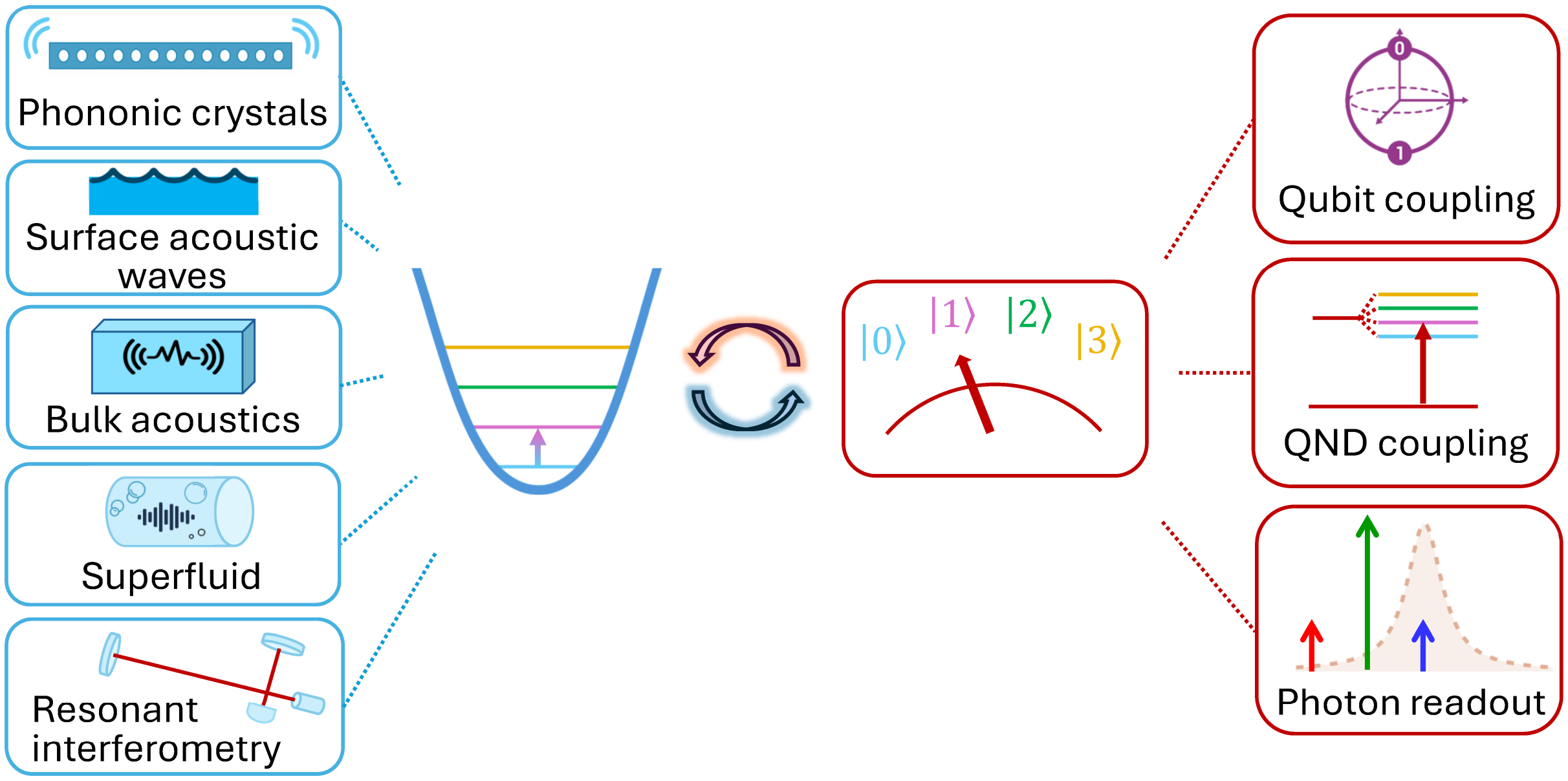

*Figure 2: Two essential ingredients of a graviton detector: a macroscopic absorber and number-resolved energy readout. A graviton can be converted into a single excitation of a harmonic mode that couples resonantly to the tidal interaction with a gravitational wave (left, blue). The resulting number state can then be read out through quantum sensing techniques (right, red), an ability recently demonstrated across a range of macroscopic systems and sensor platforms [9]. Graviton detection requires resolving individual quanta of energy in large-scale systems. Quantum acoustics provides one concrete route, but other approaches based on the same basic principles are also possible.*

### Where the gravitons come from

Using macroscopic quantum systems and monitoring their changes in energy provides a blueprint for detecting gravitons. The approach is generic based on resonant exchange of energy, and can take several forms. In addition to graviton-to-phonon conversion described above, resonant detection could also be achieved through graviton-to-photon conversion, for example in dark matter detectors [10] or with light beams [11].

But there is another challenge: a large flux of background gravitons is needed. The detection relies on a *stimulated* single graviton process, again akin to the photoelectric detection of photons. At frequencies above kilohertz, there are simply no good sources of gravitons known. But this is where many of the potential detectors, including atomic transitions, occur. This is why previous work considered thermal emission of gravitons from neutron stars, relying on the exponentially suppressed high-frequency tail of the emission, and which still required the detector to be in close proximity to the star [3].

Large quantum acoustic systems also solve this critical problem. The best graviton sources are the gravitational waves already detected by LIGO: compact binary mergers that produce a tremendous flux of gravitons at sub-kilohertz frequencies. For typical detected gravitational waves with strain amplitudes $h{\sim}10^{-22}$, the flux is on the order of

$$F \sim 10^{28}\ \frac{gravitons}{s{\cdot}m^2},$$

which compensates for the small (although no longer Planck-scale) cross-section. And these are frequencies accessible to acoustic resonators. Thus, also the graviton source problem is addressed: a macroscopic resonant detector in the frequency band of standard LIGO detections can observe the absorption of a single graviton from the passing wave.

Unfortunately, for many gravitational waves, the rapid frequency chirp, which stems from the inspiral and merger of compact binaries resulting in steadily increasing frequency, is too fast to deposit energy efficiently into such a detector. But one type of source turns out to be ideal: neutron star mergers. These last long enough for a kilogram-scale resonant acoustic system to absorb a single graviton [4]. With detectors that achieve sufficiently good phonon-number resolution or with transduction, gram-scale systems might even be sufficient.

### Solving noise through correlation

With the cross-section and graviton source addressed, the seemingly insurmountable final challenge remains background noise. Even if one can design a detector for gravitons, and the above outline shows that this is possible, how can one distinguish genuine graviton absorption from any other source of energy?

The trick is to cross-correlate the graviton detector with standard gravitational wave detections. A gravitational wave passing Earth in the right frequency band lasts only a very short time, typically less than a second. During this interval, the enormous graviton flux is present, far exceeding other potential sources. And only specific modes would be excited – those that match the spin-2 selection rules just like in an atom. The critical step is to characterize the detector when no wave is present.

Taking again the example of an acoustic resonator, the desired mode must be cooled to its ground state and its lifetime measured. Ground-state cooling of macroscopic systems has already been achieved [9], and lifetimes of minutes or more are possible, since the ground state itself is relatively mundane and not prone to extreme decoherence. The required damping and overall cooling of the whole system to about $T = 1\ mK$ are well within reach and have already been demonstrated in some systems. The noise is thus manageable enough for energy jumps and LIGO detections to be cross-correlated.

**Beyond classical wave physics**

Graviton detection thus appears to be a viable goal for experimental physics, and may well be on the horizon. The path outlined above is only one possible route. Many others can now be explored, moving the question from fantasy into the domain of quantum engineering.

But even once achieved, graviton detection would not by itself constitute a final proof of a quantum theory of gravity. This is a general feature of physics: an experiment can show consistency with a theory, but it cannot prove it uniquely. To do so in practice, one would need to rule out *all* possible alternatives, and that is rarely how physics progresses. The real question is how contrived alternative explanations would have to become before they are no longer scientifically compelling.

The photoelectric analogy offers a helpful perspective. Two opposing interpretations can be encountered, but neither captures the full significance. One, common in textbook presentations, is that the photoelectric effect by itself proved the existence of photons. The other, more common in advanced discussions of quantum optics, is that quantized matter alone is sufficient to explain the effect, and that photons are not needed at all. Both views are subtly inadequate, and the correct lesson lies between them.

The photoelectric effect, and likewise the graviton detection, *can* be described without explicitly quantizing the field. With suitable statistical assumptions, specific semi-classical models can reproduce detector clicks in quantized matter. This issue was discussed in great detail in the early days of modern quantum optics, where it became clear that ruling out semi-classical descriptions requires more refined measurements, such as antibunching.

But these are phenomenological semi-classical toy models, not ordinary classical wave physics. They are not limiting cases of full classical wave physics, but new postulated interactions. For individual particle detections, they also face a basic problem with energy conservation: energy is deposited in the detector while the classical field parameter itself remains unchanged. Conversely, assuming the validity of energy conservation, the field must lose exactly one quantum of energy, and thus graviton detection is inferred. Standard classical wave physics does not describe the extraction of discrete energy quanta E=hf, one for each detection event. It was Einstein's correct insight in 1905 that to reproduce all the hallmarks of the photoelectric observations in a consistent framework, a full quantization of both matter and light is most natural. And indeed, even today, most photon detectors rely on the photoelectric effect. It is a bona fide particle detection method.

Similarly, graviton detection would not merely be a "consistency argument", where a standard classical phenomenon is reinterpreted in a quantum language. Rather, it is a measurement of how a *single* quantum of energy is absorbed and converted from a gravitational wave, giving direct access to the single-particle process at the quantum level. This is in full analogy with how other particles are detected.

Still, graviton detection would not be final proof of the quantum nature of gravity, as it could not rule out every possible alternative explanation. In standard quantum physics, it took a range of experimental inputs to make classical physics increasingly untenable (see Figure 3). The strongest evidence to date against a fully classical description comes from recent loophole-free violations of Bell inequalities, but even those could be explained with an underlying classical nonlocal or superdeterministic model. There is only a hierarchy of increasingly stringent tests, and graviton detection would mirror the early era of this timeline.

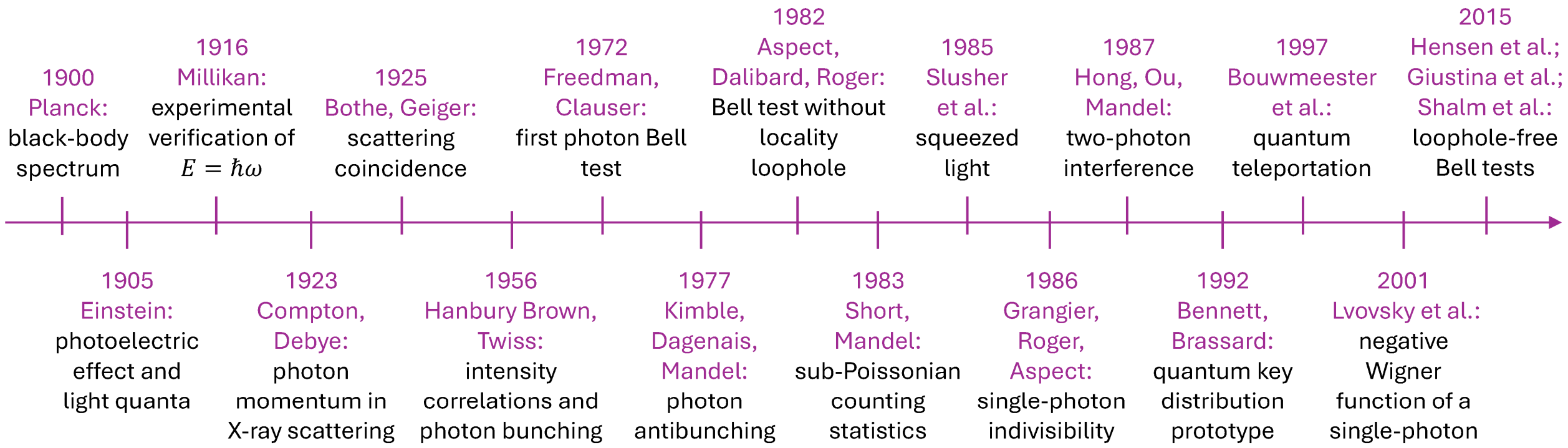

*Figure 3: Timeline of landmark experiments probing the quantum nature of light with increasing certainty. The sequence represents a hierarchy of tests of non-classicality, making classical alternatives less natural and more constrained over time. Early experiments established the particle-like behavior of light, while subsequent tests increasingly constrained alternative semiclassical models. Detecting gravitons would place gravity at the starting point of such an experimental progression, while also opening access to gravitational-wave counting statistics [12,13].*

### Towards tests of linearized quantum gravity

Detecting single gravitons would not be a one-off achievement, but would open a broader experimental program: testing the predictions of linearized quantum gravity and probing quantum features in gravitational waves. Does the graviton carry spin-2? Does it have mass? Does it follow the usual quantum relation $E = \hbar\omega$? Such properties would reveal themselves through selection rules, excitation probabilities, and the coupling to specific detector modes, much like in atomic spectroscopy for photons.

But it is now becoming clear that much more may be accessible. For example, Manikandan and Wilczek [12] recently showed that excitation probabilities involving more than one graviton can reveal whether the radiation field differs from just a coherent state. In fact, we showed that the full counting statistics of the wave through its $g^{(2)}$-function is realistically accessible in the detector even when only a few gravitons are deposited [13]. This opens the door to quantum state characterization of gravitational radiation, which may carry non-classical features, possibly arising from the inherent nonlinearity of general relativity. Graviton detection could therefore reveal a new layer of information in gravitational waves, probing both gravity and its astrophysical sources at the quantum level.

### Outlook

Gravitons are moving from being a purely mathematical prediction to becoming a realistic experimental target. But this is only part of a broader transformation in fundamental physics, in which the quantum nature of gravity is becoming experimentally accessible through the tools and insights of quantum optics and quantum information. Independent ideas for probing other quantum features of gravity have emerged over the past decade. For example, gravitational radiation could carry and imprint quantum noise, revealing a complementary signature to its particle nature [14]. Large masses may one day be placed into superposition, enabling tests of whether gravity can generate entanglement and thereby act as a quantum channel [15]. Under certain assumptions, such tests could offer insight into the quantum nature of virtual gravitons, complementing the direct detection of on-shell gravitons described here.

Taken together, a new experimental program is beginning to take shape: one in which the quantum nature of gravity can be tested step by step, with each experiment revealing a different facet of the interface between quantum physics and gravity. Dyson, while skeptical that gravitons could ever be detected, also wrote: “the great advances in science usually result from new tools rather than from new doctrines”. With the advent of quantum information technology and macroscopic quantum systems, this is precisely why quantum gravity is heading towards experiments.

**Acknowledgements**

The author acknowledges support from the W. M. Keck Foundation.